\documentclass[conference]{IEEEtran}

\usepackage{amsmath,amssymb,amsfonts}
\usepackage{algorithmic}
\usepackage{graphicx}
\usepackage{textcomp}
\usepackage{xcolor}
\usepackage{balance}
\usepackage{orcidlink}
\usepackage{eso-pic}
\hypersetup{hidelinks}

\usepackage{tikz}
\usetikzlibrary{
    shapes.misc,
    shapes.symbols,
    shapes.geometric,
    arrows.meta,
    positioning,
    backgrounds,
    fit,
    shadows,
    calc,
    patterns
}

\usepackage{booktabs}
\usepackage{array}
\usepackage{tabularx}
\usepackage{pgfplots}
\pgfplotsset{compat=1.18}

\newcommand\blfootnote[1]{%
  \begingroup
  \renewcommand\thefootnote{}\footnote{#1}%
  \addtocounter{footnote}{-1}%
  \endgroup
}

\begin{document}

\title{Pre-Crawl Prioritization and Seed Classification \\ for Large-Scale Web Archiving}

\author{
\IEEEauthorblockN{Akshith Garapati$^{*}$ \orcidlink{0009-0005-2428-0566}}
\IEEEauthorblockA{
Purdue University \\
West Lafayette, IN, USA \\
agarapat@purdue.edu 
}
\and
\IEEEauthorblockN{Sawood Alam \orcidlink{0000-0002-8267-3326}}
\IEEEauthorblockA{
Internet Archive\\
San Francisco, CA, USA \\
sawood@archive.org
}
}

\maketitle

\AddToShipoutPictureBG*{\AtPageLowerLeft{\put(\LenToUnit{\dimexpr0.5\paperwidth-0.4\textwidth},\LenToUnit{0.3in}){\parbox[b]{0.8\textwidth}{\centering\scriptsize \copyright~2025 IEEE. Personal use of this material is permitted. Permission from IEEE must be obtained for all other uses, in any current or future media, including reprinting/republishing this material for advertising or promotional purposes, creating new collective works, for resale or redistribution to servers or lists, or reuse of any copyrighted component of this work in other works. Accepted version of the paper published in the 2025 ACM/IEEE Joint Conference on Digital Libraries (JCDL), pp. 273--276, DOI: 10.1109/JCDL67857.2025.00044.}}}}

\begin{abstract}
Pre-crawl prioritization systems decide which pages to fetch first before indexing, aiming to improve crawl efficiency and coverage. Existing approaches based on page popularity and relevance are unsuited to web-archival crawlers that prioritize comprehensive coverage over selective retrieval. We explore prioritization techniques tailored to web archival operations using URL lexical analysis, dynamic domain reputation systems, and HTTP response analysis to inform crawl scheduling decisions. By incorporating predictive models that estimate content change and disappearance probabilities, we improve crawl completeness while reducing redundant visits to stable resources. Through large-scale experiments, we demonstrate that our methods improve archival effectiveness and resource utilization compared to undirected crawling.
\end{abstract}


\blfootnote{$^{*}$Work performed at the Internet Archive during affiliation.}

\section{Introduction}
Web archiving is a vital practice for safeguarding the vast and diverse information available on the internet. However, the scale and dynamism of the modern web — marked by frequent updates, link rot, and ephemeral pages — complicate the task of building comprehensive and timely archives, all the while making these very challenges what render archiving increasingly important \cite{1}.

For example, while some web pages are archived repeatedly despite no changes between versions, many short-lived pages remain unarchived due to the absence of effective ranking and selection mechanisms \cite{5}. This can limit coverage and lead to missed preservation opportunities. Moreover, these limitations reflect a fundamental difference between web archiving and search engine crawling. Search engines typically assess a page's relevance based on its content or user interaction signals, which are obtained only after the page has been fetched and processed \cite{7}. Thus, crawling is often decoupled from ranking, allowing search engines to delay more complex prioritization mechanisms until a post-crawl analysis.

In contrast, web archiving inherently couples the crawl request with content preservation. For example, once a page is requested, its content is immediately captured and stored. Since fetching a resource entails the bulk of the cost and effort of archiving, an effective pre-crawl prioritization strategy can significantly enhance crawl efficiency and coverage \cite{10}.

As a result, for archival systems, pre-crawl prioritization is critical: the decision of what to fetch next must be made predominantly using information available prior to downloading the page so as to not incur significant additional costs to the crawler’s operations. By ranking crawl requests based on predictive indicators, archival crawlers can proactively allocate limited resources to maximize the freshness and completeness of the archive, avoiding redundant retrievals of stable content and ensuring the timely capture of rapidly changing or at-risk web resources. 

Furthermore, the sheer volume of large-scale archiving can impose additional operational constraints on possible computational methods used. As a result of all these factors, web archiving systems must be carefully designed to maximize efficiency while maintaining effective coverage and data quality~\cite{11}.

In this paper, we present a novel pre-crawl prioritization framework designed specifically for large-scale web archiving. Our key contributions include: (1) a scalable URL scoring system using lexical analysis and LightGBM regression, (2) a dynamic domain reputation system that adapts to evolving site behaviors, (3) a cost-aware feature analysis module that selectively applies expensive analyses to improve system-wide accuracy, and (4) empirical validation on 25+ million URLs demonstrating 254\% improvement in score lift over heuristic baselines. 

\section{Related Work}

Web crawling systems have long prioritized pages based on quality and importance metrics. While PageRank-based approaches \cite{15} and optimal URL ordering frameworks \cite{19} show promise, they require prior knowledge of link structure and are tailored for focused crawling rather than general archival purposes. Even recent neural approaches \cite{23} require fetching page content first, making them computationally expensive for pre-crawl decisions.

Digital preservation research differentiates archival systems from search engines through unique challenges including format obsolescence and temporal coherence \cite{27}. Web archiving efforts have focused on volatility-based scheduling \cite{13}, with studies showing variable content change rates across web resources \cite{32,33}. URL-based feature extraction provides a computationally efficient approach aligned with archival system constraints, as demonstrated by lexical analysis techniques for content classification without requiring page access \cite{36}.

\section{Proposed Methodology}

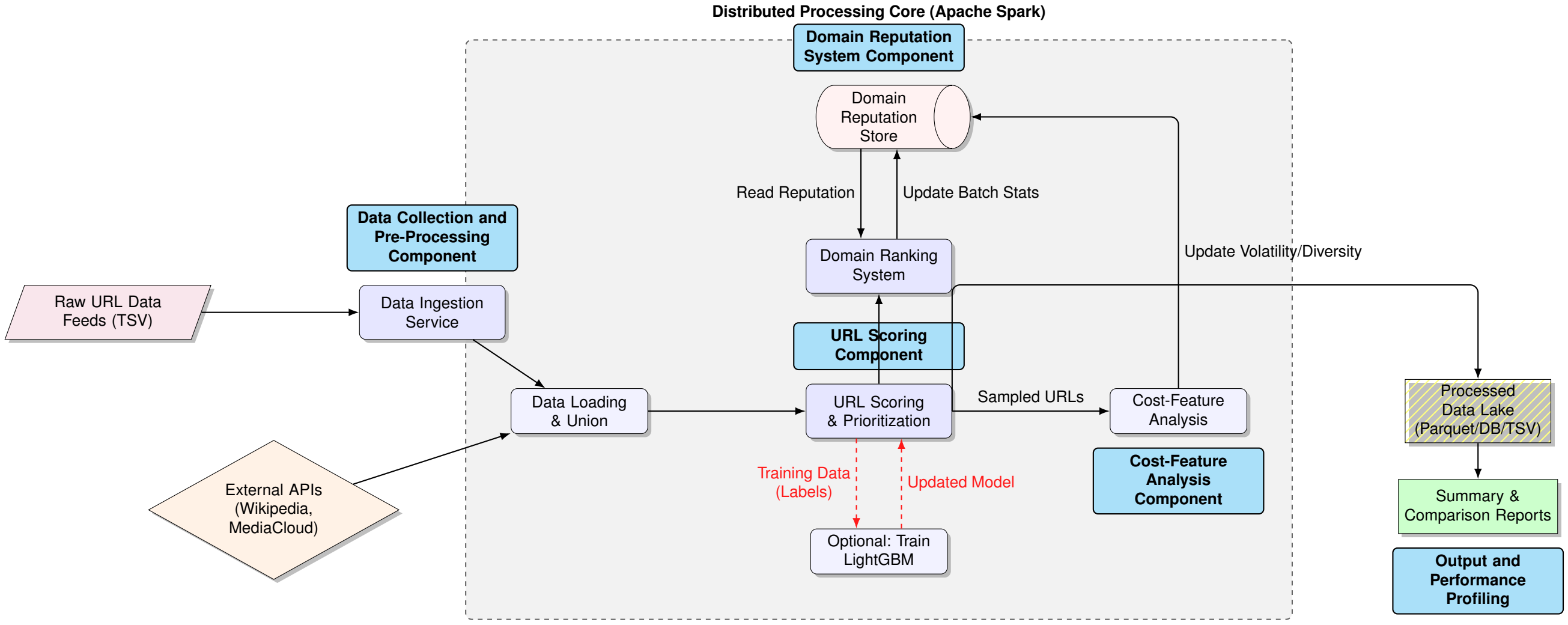
\begin{figure*}[t]
\centering
\resizebox{0.95\textwidth}{!}{%
\begin{tikzpicture}[
    font=\sffamily,
    node distance=2cm and 3.5cm,
    process/.style={rectangle, draw, fill=blue!10, rounded corners, drop shadow, minimum width=3.2cm, minimum height=1.2cm, text width=3cm, align=center},
    sub-process/.style={rectangle, draw, fill=blue!5, rounded corners, drop shadow, minimum width=3cm, minimum height=1cm, text width=2.8cm, align=center},
    database/.style={cylinder, cylinder end fill=red!10, draw, fill=red!5, shape aspect=0.6, drop shadow, minimum height=1.8cm,minimum width=1.4cm,text width=2.2cm,align=center},
    data/.style={rectangle, draw, fill=yellow!15, pattern=north east lines, pattern color=yellow!50, drop shadow, minimum width=3.2cm, minimum height=1.2cm, text width=3cm, align=center},
    io/.style={trapezium, trapezium left angle=70, trapezium right angle=110, draw, fill=purple!10, drop shadow, minimum width=3.2cm, minimum height=1.2cm, text width=2.8cm,align=center},
    api/.style={diamond, draw, fill=orange!10, aspect=1.8, drop shadow, minimum width=3cm, minimum height=1.1cm, text width=2.8cm,align=center},
    output/.style={rectangle, draw, fill=green!20, drop shadow, minimum width=3.5cm, minimum height=1.2cm, text width=3.3cm,align=center},
    container/.style={rectangle, draw=black!60, thick, dashed, fill=black!5, rounded corners, inner sep=1cm},
    arrow/.style={-Latex, thick, rounded corners=5pt},
    stepbox/.style={rectangle,draw,thick,fill=cyan!30,text=black,font=\sffamily\bfseries,rounded corners=3pt,align=center,text width=3.5cm,inner sep=4pt}
]
\node (raw_urls) [io] {Raw URL Data Feeds (TSV)};
\node (ingestion) [process, right=of raw_urls] {Data Ingestion Service};
\node (external_apis) [api, below left=3cm and 0.5cm of ingestion] {External APIs (Wikipedia, MediaCloud)};
\node (loader) [sub-process, right=of $(ingestion)!0.5!(external_apis)$] {Data Loading \& Union};
\node (scoring) [process, right=of loader] {URL Scoring \& Prioritization};
\node (ranking) [process, above=of scoring] {Domain Ranking System};
\node (datastore) [database, above=of ranking] {Domain Reputation Store};
\node (cost_feature) [sub-process, right=of scoring] {Cost-Feature Analysis};
\node (model_trainer) [sub-process, below=of scoring] {Optional: Train LightGBM};
\node (processed_data) [data, right=of cost_feature] {Processed Data Lake (Parquet/DB/TSV)};
\node (final_reports) [output, below=0.8cm of processed_data] {Summary \& Comparison Reports};
\node [stepbox, above=0.3cm of ingestion] {Data Collection and Pre-Processing Component};
\node [stepbox, above=0.3cm of scoring] {URL Scoring Component};
\node [stepbox, below=0.3cm of cost_feature] {Cost-Feature Analysis Component};
\node [stepbox, above=0.3cm of datastore] {Domain Reputation System Component};
\node [stepbox, below=0.3cm of final_reports] {Output and Performance Profiling};
\begin{pgfonlayer}{background}
    \node[container, fit=(loader) (datastore) (cost_feature) (model_trainer) (ranking)] (spark_container) {};
    \node at (spark_container.north) [above=3mm] {\textbf{Distributed Processing Core (Apache Spark)}};
\end{pgfonlayer}
\draw [arrow] (raw_urls) -- (ingestion);
\draw [arrow] (ingestion) -- (loader);
\draw [arrow] (external_apis) -- (loader);
\draw [arrow] (loader) -- (scoring);
\draw [arrow] (scoring) -- node[midway, above, sloped] {Sampled URLs} (cost_feature);
\draw [arrow] (scoring.east) -- ++(0,2.8cm) -| (processed_data.north);
\draw [arrow] (scoring.north) -- (ranking.south);
\draw [arrow] ([xshift=-0.4cm]datastore.south) -- node[midway, left] {Read Reputation} ([xshift=-0.4cm]ranking.north);
\draw [arrow] ([xshift=0.4cm]ranking.north) -- node[midway, right] {Update Batch Stats} ([xshift=0.4cm]datastore.south);
\draw [arrow] (cost_feature.north) |- node[near start, right] {Update Volatility/Diversity} (datastore.east);
\draw [arrow, dashed, red!90] ([xshift=-0.5cm]scoring.south) -- node[midway, left, align=center] {Training Data \\ (Labels)} ([xshift=-0.5cm]model_trainer.north);
\draw [arrow, dashed, red!90] ([xshift=0.5cm]model_trainer.north) -- node[midway, right] {Updated Model} ([xshift=0.5cm]scoring.south);
\draw [arrow] (processed_data) -- (final_reports);
\end{tikzpicture}
}

\caption{High-level architecture of the pre-crawl prioritization system. }
\label{fig:system_overview}
\end{figure*}

We design a system to produce ranking orders for large-scale seed URL lists used by web archiving crawlers such as Heritrix \cite{39}. Our system processes high-volume URL streams from publishers indicating recent changes, accounting for the scale and characteristics of real-world archival operations. Our system outputs ranked URL lists as TSV files to serve as ingestion feeds for crawlers with no involvement during or after the crawl process.

The code for the entire system is available under an open source license and can be found here\footnote{https://github.com/internetarchive/wbm\_seed\_stream \label{1}}. The system is structured into modular components that enable scalable URL ranking, as illustrated in Figure~\ref{fig:system_overview}: Data Collection and Pre-Processing, URL Scoring, Cost-Feature Analysis, Domain Reputation System, and Output and Performance Profiling.

The system implements core ranking logic through three interconnected modules — URL Scoring, Cost-Feature Analysis, and Domain Reputation Systems — with supporting components for data processing and performance evaluation.

\subsection{URL Scoring}

\begin{figure}[t]
\centering
\begin{tikzpicture}
\begin{axis}[
    width=0.6\columnwidth,
    height=4cm,
    font=\small,
    xlabel={Rank (i)},
    ylabel={Discount Factor},
    xmin=1, xmax=1000000,
    ymin=0, ymax=1,
    xmode=log,
    xtick={1, 10, 100, 1000, 10000, 250000, 1000000},
    xticklabels={1, 10, 100, 1K, 10K, 250K, 1M},
    legend style={
        at={(1.02,1)},
        anchor=north west,
        font=\tiny,
        cells={align=left}
    },
    grid=major,
    title={Comparison of Discount Functions for Large-Scale Ranking},
]
\addplot[
    domain=1:1000000,
    samples=200,
    color=blue,
    thick,
]
{1/(ln(x+1)/ln(2))};
\addlegendentry{Standard NDCG}
\addplot[
    domain=1:1000000,
    samples=200,
    color=red,
    thick,
]
{1 / (1 + (x / 250000)^0.533)};
\addlegendentry{Logistic Discount}
\node[circle,fill=red,inner sep=1pt] at (axis cs:1000,{1 / (1 + (1000 / 250000)^0.533)}) {};
\node[below left, font=\scriptsize, color=red] at (axis cs:1000,{1 / (1 + (1000 / 250000)^0.533)}) {(1K, 0.95)};
\node[circle,fill=red,inner sep=1pt] at (axis cs:250000,{1 / (1 + (250000 / 250000)^0.533)}) {};
\node[below left, font=\scriptsize, color=red] at (axis cs:250000,0.5) {(250K, 0.5)};
\end{axis}
\end{tikzpicture}
\caption{Comparison of the standard NDCG discount factor versus our proposed logistic discount. The logistic curve is intentionally gentle for initial ranks, aligning better with large-scale archival priorities.}
\label{fig:ndcg_decay}
\end{figure}
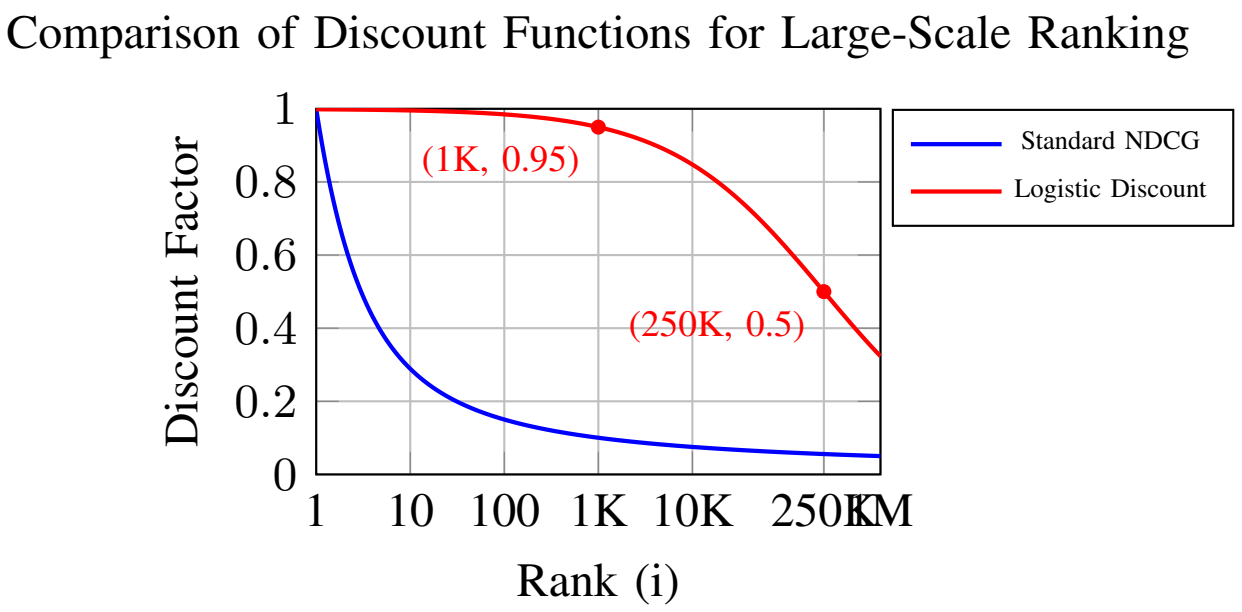

For a set of URLs $H_S = \{ u_1, u_2, \dots, u_n \}$, we produce a score function $S(u_i)$ for each URL $u_i$. The foundation of our scoring model is a feature vector $\mathbf{x}(u)$ derived exclusively from the URL string, comprising 42 different features. This lexical analysis is computationally inexpensive yet highly predictive. Formally, we define the feature vector as $\mathbf{x}(u) = [x_1(u), x_2(u), \dots, x_{42}(u)]$ where each $x_i(u)$ corresponds to a specific attribute of the URL. These features can be broadly categorized as follows:

\textbf{Structural Features}: URL length, path depth, subdomain count, and query parameters to identify spam patterns.

\textbf{Lexical Features}: Shannon entropy, vowel-to-consonant ratios, and heuristics for detecting DGAs, phishing patterns, and malicious content.

\textbf{TLD Features}: Classification of suspicious (.tk, .xyz) versus trustworthy (.gov, .edu) top-level domains.

URL lexical features are computationally efficient and effective for filtering low-quality content, with expensive features (content analysis, HTTP requests) applied selectively via the Cost-Feature Model. We frame global URL prioritization as pointwise regression using LightGBM\footnote{Model hyperparameters: learning rate 0.02, 2000 boosting iterations, max depth 8, 61 leaves per tree, L1/L2 regularization (0.1), 70\% subsampling.} to predict numerical archival priority scores, then sort URLs to generate the final ranking.

We use NDCG for evaluation, but modify the standard discount function $\frac{1}{\log_2(i+1)}$ which penalizes misplacements too aggressively for million-item crawl queues. Our logistic discount creates a gentler decay for top ranks: $d_{\text{logistic}}(i) = 1 + \left(\frac{i}{L}\right)^k$ where $L=250,000$ and $k \approx 0.533$, creating discount factors of 0.95 at rank 1,000 and 0.5 at rank 250,000 (Figure~\ref{fig:ndcg_decay}). The 250K parameter is set as an empirical measure for where crawl budgets are typically exhausted in large-scale archiving operations, but can be made adaptive to adjust the half-life to custom crawl budget constraints. The resulting $DCG_{\text{logistic}}$:

\[
DCG_{\text{logistic}}@p = \sum_{i=1}^{p} \frac{2^{rel_i} - 1}{1 + (i/L)^k}
\]

By using `LGBMRegressor', we employ a scalable, pointwise training approach to solve what is ultimately a large-scale listwise evaluation problem, with Mean Squared Error (L2 loss) as the training objective.

Training uses weak supervision with heuristic scores $S_{\text{heur}}(u) = \sum_{i=1}^{n} w_i \cdot \mathbb{I}(\text{feature } i \text{ is present in } u)$, where negative weights penalize spam indicators and positive weights reward quality signals. High-importance labels come from curated, event-driven data feeds providing signals of public interest and temporal significance.

\subsection{Domain Reputation System}

\begin{figure}[t]
\centering
\resizebox{0.6\columnwidth}{!}{%
\begin{tikzpicture}[
    font=\sffamily,
    node distance=0.8cm and 2.5cm,
    process/.style={rectangle, draw, fill=blue!10, rounded corners, drop shadow, minimum width=3.6cm, minimum height=1.3cm, text width=3.6cm, align=center},
    data/.style={rectangle, draw, fill=yellow!15, pattern=north east lines, pattern color=yellow!50, drop shadow, minimum width=3.6cm, minimum height=1.3cm, text width=3.6cm, align=center},
    database/.style={cylinder, cylinder end fill=red!10, draw, fill=red!5, shape aspect=0.6, drop shadow, minimum width=3.2cm, minimum height=1.6cm, text width=3.2cm, align=center},
    api/.style={diamond, draw, fill=orange!10, aspect=1.8, drop shadow, minimum width=3.0cm, minimum height=1cm, text width=3.0cm, align=center},
    io/.style={trapezium, trapezium left angle=70, trapezium right angle=110, draw, fill=purple!10, drop shadow, minimum width=3.2cm, minimum height=1.2cm, text width=3.2cm, align=center},
    output/.style={rectangle, draw, fill=green!20, drop shadow, minimum width=3.6cm, minimum height=1.3cm, text width=3.6cm, align=center},
    trigger/.style={ellipse, draw, fill=cyan!10, drop shadow, minimum width=3.0cm, minimum height=1.2cm, text width=3.0cm, align=center},
    ext_api/.style={ellipse, draw, fill=gray!10, drop shadow, minimum width=2.5cm, text width=2.5cm, align=center},
    arrow/.style={-Latex, thick, rounded corners=5pt}
]
    \node (driver) [process] {Batch Coordinator (Spark Driver)};
    \node (augmentation) [process, below=of driver] {Data Loading \& Augmentation};
    \node (pass1) [process, below=of augmentation] {Pass 1: Provisional Scoring (with Historical Reputation)};
    \node (aggregate) [process, below=of pass1] {Aggregate Batch-Wide Domain Stats};
    \node (pass2) [process, below=of aggregate] {Pass 2: Rescore URLs (with Batch Stats)};
    \node (final_df) [data, below=of pass2] {Final Scored URLs (DataFrame)};
    \node (db) [database, right=4cm of driver] {Domain Reputation Store};
    \node (output_formats) [output, below=6.0cm of db] {Output (Parquet/DB/TSV)};
    \node (sampling) [process, below=of final_df] {Optional: Sample URLs (1 per domain)};
    \node (feature_analysis) [process, below=of sampling] {Deep Feature Analysis};
    \node (cdx_api) [ext_api, right=of feature_analysis] {CDX API};
    \node (http_req) [ext_api, below=of cdx_api] {Live HTTP Requests};
    \draw [arrow] (driver.south) -> (augmentation.north);
    \draw [arrow] (augmentation) -> (pass1);
    \draw [arrow] (pass1) -> (aggregate);
    \draw [arrow] (aggregate) -> (pass2);
    \draw [arrow] (pass2) -> (final_df);
    \path (driver.east) -- (db.west) coordinate[pos=0.5] (gap);
    \coordinate (corner) at (gap |- pass1.east);
    \draw [arrow] (db.west) -- (gap) |- (pass1.east);
    \path (gap) -- (corner) node [midway, left=4pt, rotate=90, anchor=south] {Read Historical Reputation};
    \draw [arrow] (aggregate.east) -| (db.south) node [midway, above, rotate=90, pos=0.75] {Update with EMA};
    \draw [arrow] (final_df.east) -| (output_formats.south);
    \draw [arrow, dashed] (final_df) -> (sampling);
    \draw [arrow, dashed] (sampling) -> (feature_analysis);
    \draw [arrow, dashed] (feature_analysis) -> (cdx_api);
    \draw [arrow, dashed] (feature_analysis) -> (http_req);
    \draw [arrow, dashed] (feature_analysis.west) -- ++(-0.4, 0) |- (final_df.west) node [pos=0.5, left] {Enrich};
    \begin{pgfonlayer}{background}
        \draw [arrow, dashed] (feature_analysis.east) .. controls ++(2,0) and ++(-3,0) .. (db.west) node [midway, above, sloped, pos=0.55] {Update Volatility};
    \end{pgfonlayer}
\end{tikzpicture}
}
\caption{Workflow of the Domain Reputation System.}
\label{fig:domain_reputation}
\end{figure}
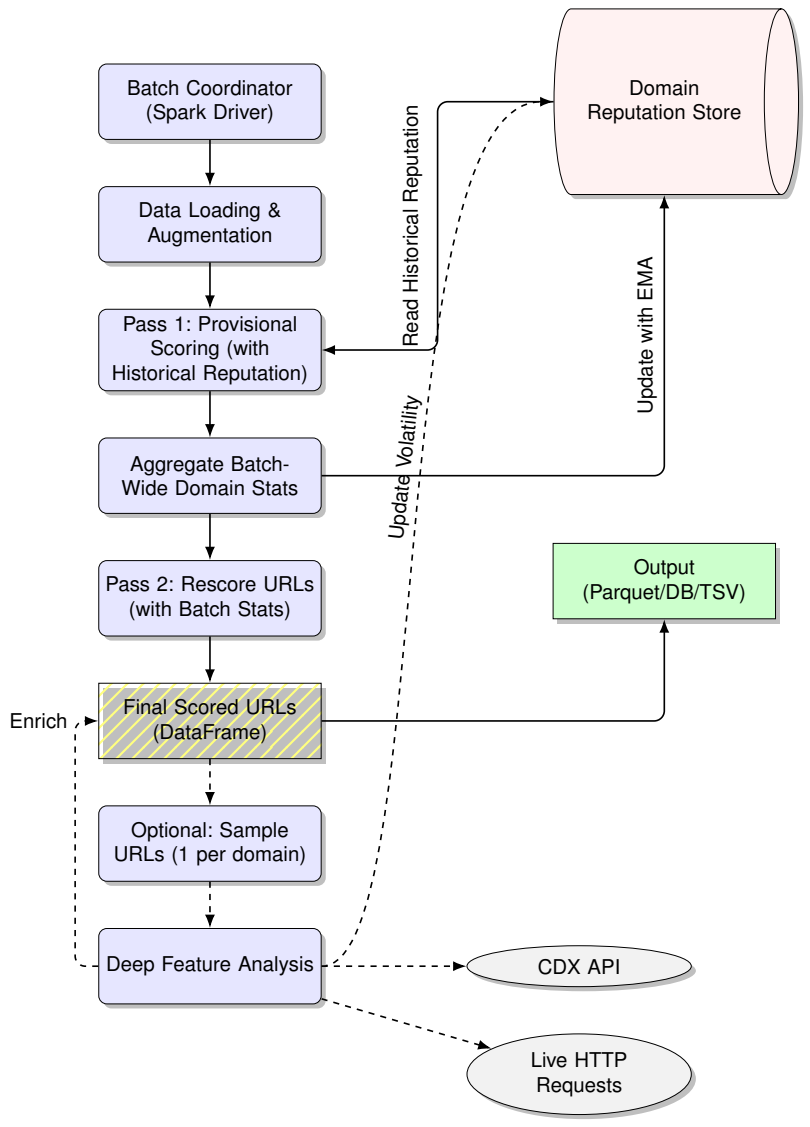

The Domain Reputation System maintains dynamic profiles for observed domains, aggregating URL observations to produce reputation, volatility, and diversity scores that inform URL scoring decisions (Figure~\ref{fig:domain_reputation}). The state of each domain $j$ is defined by a vector of learned attributes stored in a persistent database, including its reputation score $s_j$, content volatility $v_j$, and content diversity $d_j$. These attributes are updated with each processed batch of URLs.

The core reputation score, $s_j$, is updated using an Exponential Moving Average (EMA), which gives more weight to recent observations while retaining memory of past behavior. Given a domain's reputation score $s_j^{(t-1)}$ from the previous state and the mean score $\bar{s}_j^{(t)}$ of all URLs from that domain in the current batch, the new score $s_j^{(t)}$ is calculated as:

$$s_j^{(t)} = (1 - \alpha) s_j^{(t-1)} + \alpha \bar{s}_j^{(t)}$$

where $\alpha$ is the smoothing factor, set to $\alpha=0.1$ to ensure that the reputation remains stable and is not overly influenced by small, anomalous batches.

Content Volatility score $v_j$ (from HTTP header changes) determines Predicted Revisit Interval:

$$T_{\text{revisit}}(j) = \max \left( T_{\min}, T_{\text{base}} \cdot \left(1 - \frac{\min(v_j, v_{\max})}{v_{\max}}\right) \right)$$

where $T_{\text{base}}$ is the maximum revisit interval (e.g., 720 hours), $T_{\min}$ is the minimum interval (1 hour), and $v_{\max}$ is a cap on the volatility influence.

Content Diversity score $d_j$ distinguishes single-topic domains from multi-topic platforms using URL score variance and path segment cardinality, dynamically adjusting domain reputation weight in $S_\text{heur}(u)$.

\subsection{Cost-Feature Module}

\begin{figure}[t]
\centering
\resizebox{0.5\columnwidth}{!}{%
\begin{tikzpicture}[
    node distance=0.75cm and 1.5cm,
    process/.style={rectangle, rounded corners, draw=blue!80, fill=blue!10, thick, minimum height=1cm, minimum width=4.5cm, align=center, drop shadow},
    decision/.style={diamond, aspect=1.5, draw=orange!80, fill=orange!10, thick, minimum size=1cm, align=center, drop shadow},
    io/.style={trapezium, trapezium left angle=70, trapezium right angle=110, draw=gray!80, fill=gray!10, thick, minimum height=1cm, align=center, drop shadow},
    database/.style={cylinder, shape border rotate=90, aspect=0.25, draw=purple!80, fill=purple!10, thick, minimum height=1.5cm, align=center, drop shadow},
    connector/.style={-Stealth, thick, draw=black!70},
    path_yes/.style={connector, draw=green!50!black},
    path_no/.style={connector, draw=red!70!black}
]
\node[io] (input) {Primary Scores Table};
\node[process, below=of input] (create_sample) {Create URL Sample \\ (Per-domain, de-skewed, limited size)};
\node[process, below=of create_sample] (cheap_analysis) {\textbf{Lower Cost:} Cheap Analysis \\ (HTTP Request, CDX API)};
\node[decision, below=of cheap_analysis] (is_accessible) {URL Accessible? \\ (Status 200 OK)};
\node[process, below=of is_accessible] (expensive_analysis) {\textbf{Higher Cost:} Expensive Analysis \\ (Full Content, Outlinks)};
\node[process, below=of expensive_analysis] (calculate_score) {Calculate Enhanced Metrics \\ (Score Delta, Volatility Signals)};
\node[process, below=of calculate_score] (aggregate_results) {Process \& Aggregate Results};
\node[database, below=of aggregate_results] (db_update) {Batch Update Domain Statistics};

\draw[connector] (input) -- (create_sample);
\draw[connector] (create_sample) -- (cheap_analysis);
\draw[connector] (cheap_analysis) -- (is_accessible);
\draw[path_yes] (is_accessible.south) -- node[left, midway] {Yes} (expensive_analysis.north);
\draw[connector] (expensive_analysis) -- (calculate_score);
\draw[path_no] (is_accessible.east) -- ++(2.8,0) |- (aggregate_results.east) node[pos=0.25, right] {No};
\draw[connector] (calculate_score) -- (aggregate_results);
\draw[connector] (aggregate_results) -- (db_update);
\end{tikzpicture}
}
\caption{Workflow of the Cost-Feature Analysis Module.}
\label{fig:cost_feature}
\end{figure}
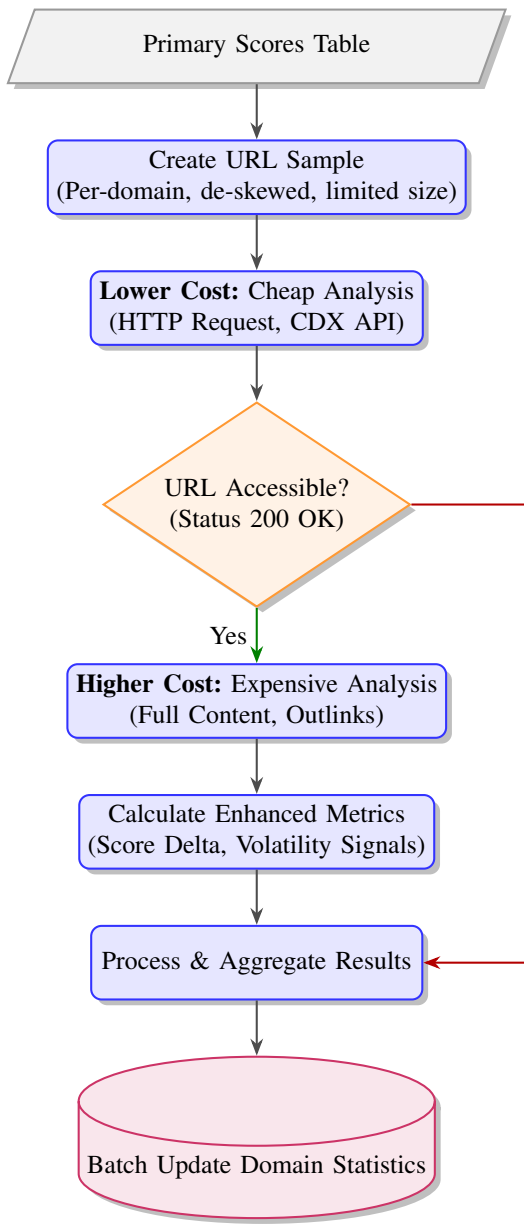

The Cost-Feature Module applies expensive analyses (HTTP requests, content analysis, CDX queries) to strategically sampled URLs, updating the Domain Reputation System to benefit the primary lexical model. We use budget-constrained, domain-stratified sampling.

Given an incoming batch of URLs $H_S$, the selection process is as follows: (1)
Partition the set $H_S$ by domain, creating disjoint subsets $\{H_{S,j}\}$ for each unique domain $j$. (2) From each non-empty subset $H_{S,j}$, select exactly one URL, $u_j^*$, uniformly at random. This creates a candidate set $H_S^* = \{u_j^*\}$ where each domain is represented once. (3) To adhere to a fixed computational budget, if the size of the candidate set $|H_S^*|$ exceeds a predefined maximum $K$, a final random sample of size $K$ is drawn from $H_S^*$. In our implementation, $K$ is set to 6,500.

Sampled URLs undergo deeper analysis computing $\mathbf{z}(u)$ consisting of the 4 aforementioned features. Enhanced scores $S'(u) = S(u) + \Delta(\mathbf{z}(u))$ and domain-level aggregates update the domain reputation table, creating a feedback loop where expensive analysis benefits future URL scoring (Figure~\ref{fig:cost_feature}).

\section{Experiments and Results}

We evaluated our pre-crawl prioritization framework by comparing two system versions using approximately 25 million URLs from the IndexNow \cite{41} protocol and 500,000 URLs from MediaCloud \cite{43} collected on June 18, 2025. The LightGBM model was trained on a similar-sized dataset of 25.5 million URLs from an earlier time period to ensure no temporal overlap between training and test data. We compared an unoptimized heuristic-based system using TLD priority lists, good/bad domain lists, and fixed empirical weights for URL features against our complete system incorporating the trained LightGBM regressor.

\begin{table}[t]
\centering
\caption{Comparison of System Performance}
\label{tab:system_comparison}
\renewcommand{\arraystretch}{0.9}
\begin{tabular}{lcc}
\toprule
\textbf{Metric} & \textbf{Unoptimized} & \textbf{Optimized} \\
\midrule
Total URLs & 25.5M & 25.5M \\
Score Lift & -59.77\% & +254.75\% \\
Spam URLs & 2,237,397 & 529,059 \\
Clean Rate & 91.02\% & 99.63\% \\
Score Std. Dev. & 0.2371 & 0.6578 \\
Avg. Confidence & 0.9951 & 0.6005 \\
\bottomrule
\end{tabular}
\end{table}

The results demonstrate significant improvements with the optimized system. Most notably, the score lift, defined as the percentage change in ranking position of high-quality URLs (from event-driven feeds) compared to random ordering, improved from -59.77\% to +254.75\%, indicating the machine learning approach correctly prioritizes high-value content while the heuristic system paradoxically scored valuable URLs lower than average content.

The optimized system shows superior spam filtering with 76\% fewer false positives, achieving a 99.63\% clean data rate. The wider score distribution (standard deviation 0.6578 vs 0.2371) enables more nuanced prioritization decisions, while moderate confidence levels suggest realistic uncertainty estimation rather than overconfident predictions. This drop in average confidence reflects the model's appropriate uncertainty on a large test set of diverse URLs, indicating well-calibrated estimates that distinguish genuine uncertainty from known patterns.

\vspace{-0.025in}

\section{Conclusion and Future Work}
We presented a pre-crawl prioritization framework combining lexical URL analysis, dynamic domain reputation, and cost-aware feature analysis. Our experiments demonstrate improved efficiency and effectiveness by directing resources toward volatile, at-risk, and high-value content, highlighting pre-crawl decision-making's potential for web archiving.

Future work includes tighter integration between prioritization and crawling strategies, using our scores to modulate runtime crawl parameters enabling more responsive archival coverage.

\vspace{-0.03in}

\section{Acknowledgements}

This work was conducted during Google Summer of Code 2025, supported by the resources and infrastructure of the Internet Archive. We thank Will Howes for his guidance and support throughout this research. We also acknowledge IndexNow and Media Cloud for providing access to the data essential to this work.

\end{document}